\documentclass[aps,superscriptaddress,amsmath,amssymb,preprintnumbers,showpacs,11pt]{revtex4}

\usepackage{txfonts}

\usepackage{amssymb}

\usepackage{indentfirst}

\usepackage{bm}

\usepackage{epsfig}

\usepackage{epsf}

\usepackage{graphicx}

\newcommand{\bT}{\boldsymbol b_T}
\newcommand{\DT}{\boldsymbol \Delta_T}
\newcommand{\nT}{\boldsymbol 0_T}
\newcommand{\kT}{\boldsymbol k_T}

\newcommand{\pT}{\boldsymbol p_T}

\newcommand{\ppT}{{\boldsymbol p\,}'_{\!T}}

\def\beq{\begin{equation}}
\def\eeq{\end{equation}}

\newcommand{\be}{\begin{equation}}
\newcommand{\ee}{\end{equation}}
\newcommand{\ba}{\begin{eqnarray}}
\newcommand{\ea}{\end{eqnarray}}

\begin{document}

\title{Orbital structure of quarks inside the nucleon in the light-cone diquark model}

\author{Zhun Lu}
\affiliation{Department of Physics, Southeast University, Nanjing
211189, China}
\affiliation{Departamento de F\'\i sica, Universidad T\'ecnica
Federico Santa Mar\'\i a, and Centro Cient\'\i fico-Tecnol\'ogico de
Valpara\'\i so Casilla 110-V, Valpara\'\i so, Chile}
\author{Ivan Schmidt}
\affiliation{Departamento de F\'\i sica, Universidad T\'ecnica
Federico Santa Mar\'\i a, and Centro Cient\'\i fico-Tecnol\'ogico de
Valpara\'\i so Casilla 110-V, Valpara\'\i so, Chile}

\begin{abstract}
We study the orbital angular momentum structure of the quarks inside
the proton. By employing the light-cone diquark model and the
overlap representation formalism, we calculate the chiral-even
generalized parton distribution functions (GPDs)
$H_q(x,\xi,\Delta^2)$, $\widetilde{H}_q(x,\xi,\Delta^2)$ and
$E_q(x,\xi,\Delta^2)$ at zero skewedness for $q=u$ and $d$
quarks. In our model $E_u$ and $E_d$ have opposite sign with similar
size. Those GPDs are applied to calculate the orbital angular
momentum (OAM) distributions, showing that $L_u(x)$ is positive, while
$L_d(x)$ is consistent with zero compared with $L_u(x)$. We
introduce the impact parameter dependence of the quark OAM distribution. It
describes the position space distribution
of the quark orbital angular momentum at given $x$. We found that
the impact parameter dependence of the quark OAM
distribution is axially symmetric in the light-cone diquark model.

\end{abstract}

\pacs{12.39.Ki, 13.88.+e, 14.20.Dh}

\maketitle

\section{introduction}

understanding the spin structure of the nucleon is one of the
most important challenges in hadron physics. The naive picture that
the nucleon spin is provided totally by the spin of its three
valence quark was proved to be wrong by the experimental
measurements. The EMC result~\cite{emc} indicates that a large
fraction of the nucleon spin is carried by other sources of angular
momentum. There have been many attempts to explain the EMC result
from the fundamental theory. Besides the angular momentum of the
gluon, the quark orbital angular momentum (OAM)~\cite{Sehgal:1974rz}
is believed to provide a substantial part of the nucleon spin.
In the last two decades the theoretical description of the quark OAM distribution
has been established~\cite{orbital,ji97,hagler98,kundu99,Ma:1998ar}.
It has been shown by Ji that the quark angular momentum can be
separated into~\cite{ji97} the usual quark helicity and a
gauge-invariant orbital contributions $L_q$. One of the advantage of
this decomposition is that $L_q$ is related to generalized parton
distributions
(GPDs)~\cite{Mueller:1998fv,Ji:1996nm,Radyushkin:1997ki,diehl03,Belitsky:2005qn,Boffi:2007yc},
the experimental observables that enter the descriptions of hard
exclusive processes, such as deeply virtual Compton
processes~\cite{Radyushkin:1996nd,Ji:1996nm} and meson exclusive
production~\cite{Polyakov:1998ze,Collins:1996fb}.

Moreover, recently it has been found that the quark OAM plays an
essential role through spin-orbit correlations in some novel
phenomena that appear in the physics of single spin asymmetries,
among which a particular transverse momentum distribution
(TMD)~\cite{Mulders:1995dh,Boer:1997nt}---- Sivers
function~\cite{Sivers:1989cc,Sivers:1990fh}----has attracted a lot
of interest, since it is an essential piece in our understanding of
the single spin asymmetries (SSA) observed in semi-inclusive deeply
inelastic scattering (SIDIS). These SSAs have been measured recently
by both the HERMES~\cite{Airapetian:2004tw,hermes05} and
COMPASS~\cite{compass,compass06} Collaborations. An interesting
observation is that there is a quantitative
relation~\cite{amm,Burkardt:2005km,ls07} between the Sivers function
$f_{1T}^{\bot q}$ and the GPD $E^q$, although it is obtained in a
model dependent way, suggesting that similar underlying physics
plays a role for nonzero $f_{1T}^{\bot q}$ and $E^q$. Similar relations
have been obtained between Boer-Muldes functions and chiral-odd quark
GPDs~\cite{Diehl:2005jf,Gockeler:2006zu}. A complete study on the
relations between the GPDs and TMDs has been presented in
\cite{gpdtmd}, which becomes more transparent through the conception of
general parton correlation
functions~\cite{Meissner:2008ay,Meissner:2009ww}. The relations
between GPDs and TMDs are more
intuitive~\cite{Burkardt:2002ks,Burkardt:2003uw} if we interpret
GPDs in the transverse position (impact parameter)
space~\cite{Burkardt:2000za,Burkardt:2002hr,Diehl:2002he,Burkardt:2003je}.
Of particular interest is the case of zero skewedness ($\xi = 0$),
where a density interpretation of GPDs in impact parameter space may
be obtained~\cite{Burkardt:2000za}. In particular this
interpretation allows one to study a three-dimensional picture of
the nucleon.

In this paper, we study the orbital angular momentum structure of
the quarks inside the proton in a light-cone diquark model. In this
model the light-cone wave function of the proton can be obtained. It is
then convenient to express the physical observables in the overlap
representation formalism~\cite{Brodsky:2000xy,Diehl:2000xz}. We
calculate the chiral-even generalized parton distribution functions
(GPDs) $H_q(x,\xi,\Delta^2)$, $\widetilde{H}_q(x,\xi,\Delta^2)$ and
$E_q(x,\xi,\Delta^2)$ at the zero skewedness for $q=u$ and $d$. We
found that $E_u$ and $E_d$ have opposite sign with similar size in
this model. The GPDs are applied to calculate the quark OAM  distributions,
showing that $L_u(x)$ is positive, while
$L_d(x)$ is consistent with zero compared with $L_u(x)$, and the net
OAM of the $u$ and $d$ quarks is positive. We
also introduce the impact parameter dependence of quark OAM distribution.
It describes the position space
distribution of the quark OAM at given $x$. We
found that the impact parameter dependence of quark OAM distribution is axially symmetric in the light-cone diquark
model.

The manuscript is organized in the following way. In Section.~II we
review the GPDs and their connections with quark orbital
angular momentum, in Section III, we present the calculation of
chiral-even GPDs from the light-cone diquark model, by applying the
overlap representation formalism. We also show the calculation of
the quark OAM in the same approach. In Section IV we introduce
the impact parameter dependence of quark OAM distribution and present results of the position space distribution
for orbiting $u$ quark, from the light-cone diquark model. We
summarize our paper in Section V.

\section{systematics of generalized parton distributions
and the orbital angular momentum}

GPDs are introduced to describe the exclusive process in which the
momenta of the incoming and outgoing nucleon in the symmetric frame
are given by
\begin{equation}
 p=P+\tfrac{1}{2}\Delta \,, \qquad
 p'=P-\tfrac{1}{2}\Delta \,,
\end{equation}
and satisfy $p^2 = p'^2 = M^2$, with $M$ denoting the nucleon mass.
The GPDs depend on the following variables
\begin{equation}
 x=\frac{k^+}{P^+} \,, \qquad
 \xi=-\frac{\Delta^+}{2P^+} \,, \qquad
 t=\Delta^2 \,,
\end{equation}
where the light-cone coordinates are defined by
\begin{equation}
 a^\pm=(a^0\pm a^3) \,, \qquad
 \vec{a}_T=(a^1,a^2)
\end{equation}
for a generic 4-vector $a$. In a physical process the so-called
skewness $\xi$ and the momentum transfer $t$ to the nucleon are
fixed by the external kinematics, whereas $x$ is typically an
integration variable.

The chiral-even GPDs $H_q$, $E_q$ and ${\widetilde H}_q$,
${\widetilde E}_q$ for quarks are defined through matrix elements of
the bilinear vector and axial vector currents on the light-cone:
\begin{eqnarray}
\lefteqn{ \int\frac{d y^-}{8\pi}\;e^{ix P^+y^-/2}\; \langle p' |
\bar\psi(0)\,\gamma^+\,\psi(y)\,|p\rangle \Big|_{y^+=0, y_\perp=0} }
\label{spd-def} \\
&=& {1\over 2 P^+}\ {\bar U}(p') \left( \,\ {\gamma^+} H(x,\xi,t)
 + \ {i\sigma^{+\mu}\Delta_\mu\over 2M}
E(x,\xi,t)\, \right) U(p)\ ,
\nonumber\\
\lefteqn{ \int\frac{d y^-}{8\pi}\;e^{ix P^+y^-/2}\; \langle p' |
\bar\psi(0)\,\gamma^+\gamma_5\,\psi(y)\,|p\rangle \Big|_{y^+=0,
y_\perp=0} }
\\
&=& {1\over 2 P^+}\ {\bar U}(p') \left( \, \
{\gamma^+\gamma_5}{\widetilde H}(x,\xi,t)
 +\ {\Delta^+ {\gamma_5}\over 2M}\,
{\widetilde E}(x,\xi,t)\, \right ) U(p)\ . \nonumber
\end{eqnarray}

An important implication of GPDs is that they are related to the
OAM (OAM ) of the quark, which is expected to
provide essential contribution to the total spin of the nucleon.
Here we follow the decomposition of the nucleon spin introduced by
Ji~\cite{ji97}:
\begin{equation}
J^z= J_q^z+ J_g^z= \frac{1}{2}\sum_q \Delta q + \sum_q L_q^z +
J_g^z=\frac{1}{2} ,
\end{equation}
where $ \Delta q$, $L_q^z$ and $J_g^z$
 denote the quark spin, quark OAM  and gluon angular momentum, which
 comes from the expectation value of the operator
\begin{equation}
M^{0xy}=\frac{1}{2}\sum_q  \psi_q^\dag \Sigma^z \psi_q + \sum_q
\psi_ q^\dag (\vec r \times i \vec D)^z \psi_q + [\vec r \times
(\vec E \times \vec B)]^z, \label{J-decompose}
\end{equation}
Note that in literature~\cite{orbital,kundu99,Burkardt:2008ua} there are some other ways to decompose the
nucleon spin. The advantage of the decomposition of $J_q$ to $\Delta
q$ and $L_q^z$ in (\ref{J-decompose}) is that it ensures the gauge
invariance of the operators. There has been also discussion that wether the
gluon angular momentum can be further decomposed gauge-invariantly.
In this work we will not consider the gluon contribution.

\begin{figure}
\begin{center}
\scalebox{0.34}{\includegraphics*[22pt,49pt][609pt,524pt]{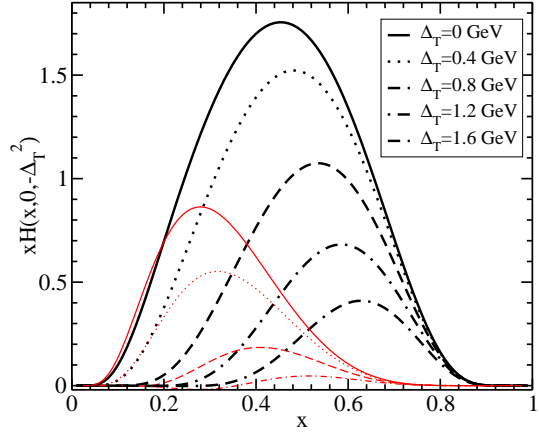}}
\caption{\small The generalized parton distributions
$H_u(x,0,\Delta_T^2)$
 and $H_d(x,0,\Delta_T^2)$  for the proton in the light-cone diquark model as functions of x
 for different values of $\Delta_T$.}\label{xH-ud}
\end{center}
\end{figure}

The quark OAM  distribution $L_q(x)$ can then be defined as the
expectation value of operator
\begin{eqnarray}
\hat{O}_{L}=\int d\eta \textrm{e}^{-i x P^+\eta }\psi_q^\dag (\vec r
\times i \vec D)^z \psi_q, \label{oam-operator}
\end{eqnarray}
between the proton state $|P \,S\rangle$:
\begin{equation}
L_q^z(x)= \left\langle P \,S \left | \hat{O}_{L}  \right |P\,S
\right\rangle
\end{equation}
The quark OAM  distribution can be obtained
from~\cite{ji97,Hoodbhoy:1998yb}
\begin{eqnarray}
L_q^z(x)= \frac{1}{2}\left \{x\left[H_q(x,0,0)+E_q(x,0,0)\right
]-\widetilde{H}_ q(x,0,0)\right \},\label{lqz}
\end{eqnarray}
where $H_q(x,0,0)$, $\widetilde{H}_q(x,0,0)$ and $E_q(x,0,0)$ are
the forward limits of GPDs. Furthermore, the former two are the
unpolarized and helicity distributions for the nucleon,
respectively,
\begin{eqnarray}
q(x)= H_q(x,0,0), ~~~~ \Delta q(x)=\widetilde{H}_ q(x,0,0),\label{lqz}
\end{eqnarray}
and $E_q(x,0,0)$ is related to the anomalous magnetic momentum of
the nucleon in the following way:
\begin{eqnarray}
\int_0^1 dx E_q(x,0,0) =\kappa_q,
\end{eqnarray}
where $\kappa_q$ is the contribution of quark flavor $q$ to the
nucleon anomalous magnetic momentum.

\section{GPDs in the light-cone diquark model from the overlap representation formalism}

\begin{figure}
\begin{center}
\scalebox{0.35}{\includegraphics*[15pt,46pt][593pt,508pt]{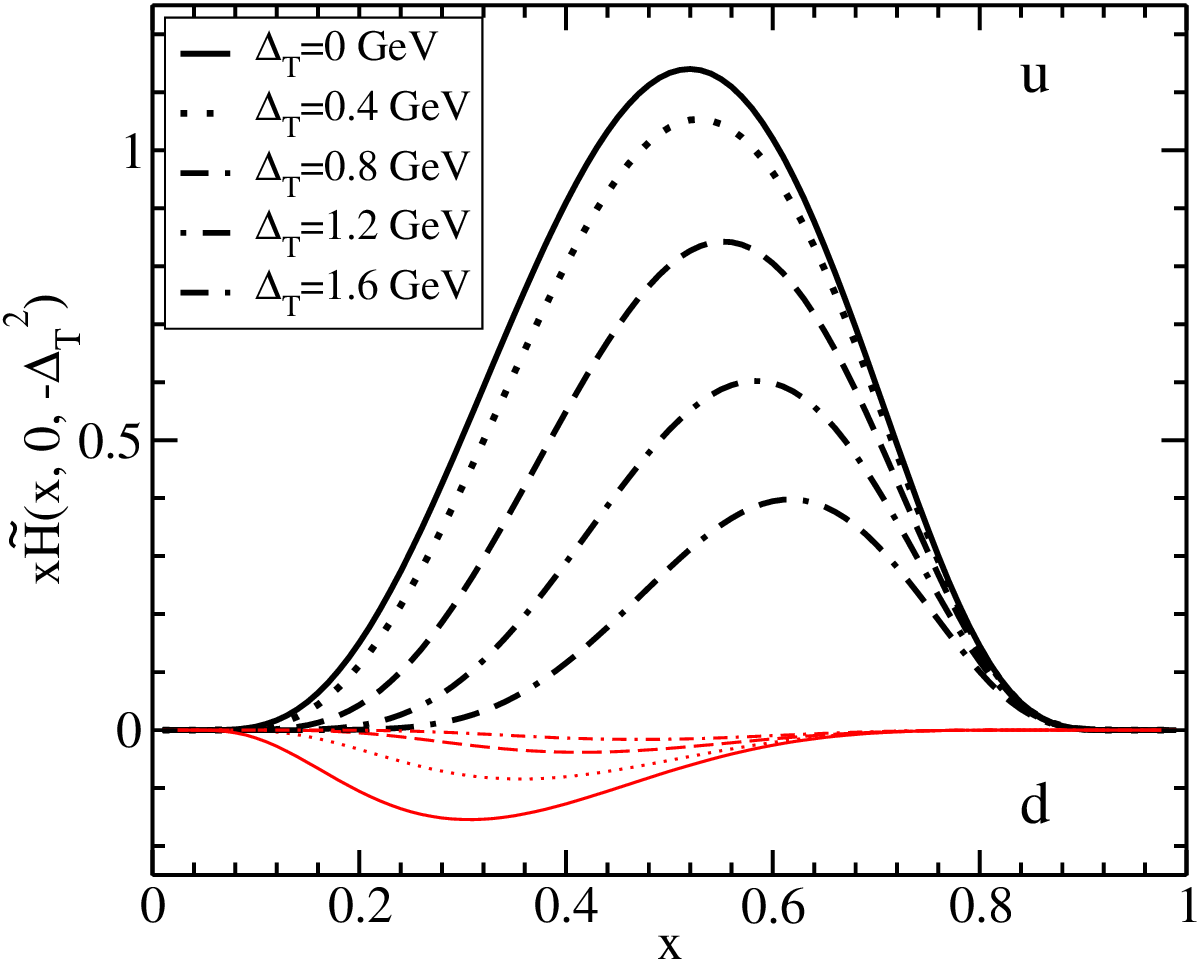}}
\caption{\small The generalized parton distributions
$\widetilde{H}_u(x,0,\Delta_T^2)$ and
$\widetilde{H}_d(x,0,\Delta_T^2)$ for the proton in the light-cone
diquark model as functions of x for different values of
$\Delta_T$.}\label{Htidle-ud}
\end{center}
\end{figure}

In this section we present the calculation of the GPDs in the
light-cone diquark model from the overlap representation formalism.
The proton wave function with helicity $\Uparrow,\Downarrow$ in the SU(6) quark-diquark
model~\cite{ref:qdq,Ma,Ma:2002ir} in the instant form is written
 as
 \begin{equation}
 \Psi^{\Uparrow,\Downarrow}(qD)=\frac{1}{\sqrt{2}} \varphi_V|qV\rangle^{\Uparrow,\Downarrow}
 +\frac{1}{\sqrt{2}} \varphi_S|qS\rangle^{\Uparrow,\Downarrow},\label{eq:qdq}
\end{equation}
where $D=V,S$ denotes the vector diquark and scalar diquark,
respectively.
The
\begin{eqnarray}
|qV\rangle^{\Uparrow,\Downarrow}&=&\pm\frac{1}{3}
[V_0(ud)u^{\uparrow,\downarrow}-
\sqrt{2}V_{\pm1}(ud)u^{\downarrow,\uparrow}\nonumber\\
 &&-\sqrt{2}V_0(uu)d^{\uparrow,\downarrow}+
2V_{\pm1}(uu)d^{\downarrow,\uparrow}];\nonumber\\
|qS\rangle^{\Uparrow,\Downarrow}&=&S(ud)u^{\uparrow,\downarrow},
\end{eqnarray}

The spin part of the light-cone wave function of the proton can be
obtained from the instant form of the wave
function by a Melosh rotation. For a spin-$\frac{1}{2}$ particle, the Melosh
transformations are known to be \cite{ref:melosh}
\begin{eqnarray}\label{eqn:melosh}
\chi^{\uparrow}_T &=& \omega \left[\left(k^+ + m_q\right) \chi^{\uparrow}_F
- k^R \chi^{\downarrow}_F\right],
\nonumber \\
\chi^{\downarrow}_T &=& \omega \left[\left(k^+ + m_q\right)
\chi^{\downarrow}_F + k^L \chi^{\uparrow}_F\right],
\end{eqnarray}
where $\chi_T$ and $\chi_F$ are instant and light-cone spinors
respectively,
$\omega=\left[2k^+\left(k^0+m_q\right)\right]^{-\frac{1}{2}}$,
$k^{R,L}=k^1\pm i k^2$, and $m_q$ is the quark mass. In this work, for
simplicity we treat the diquark as a point particle. The scalar
diquark does not transform, since it has zero spin. For the spin-$1$
vector diquark, the Melosh transformations are given by
\cite{ref:as}
\begin{eqnarray}\label{eqn:melosh1}
V^1_T &=& \omega_V^2 \left[\left(k_V^+ +\lambda_V\right)^2 V^1_F
-\sqrt{2}\left(k_V^+ +\lambda_V\right)k_V^R V^0_F + {k_V^R}^2 V^{-1}_F\right],\nonumber\\
V^0_T &=& \omega_V^2 \left[\sqrt{2}\left(k_V^+ +\lambda_V\right)k_V^L V^1_F +
2\left(\left(k_V^0+\lambda_V\right)k_V^+-k_V^R k_V^L\right)V^0_F \right .
\nonumber\\
&&-\sqrt{2} \left .\left(k_V^+
+\lambda_V\right)k_V^R V^{-1}_F\right], \\
V^{-1}_T &=& \omega_V^2 \left[{k_V^L}^2 V^1_F+\sqrt{2}\left(k_V^+ +\lambda_V\right)k_V^L
V^0_F + \left(k_V^+ +\lambda_V\right)^2V^{-1}_F\right]\nonumber.
\end{eqnarray}
Here, $\lambda_V$ denotes the mass of the diquark, $V_T$ and $V_F$ are the instant and light-cone spin-$1$
particle respectively, which are constructed within the
Weinberg-Soper formalism \cite{weinberg64}.

After some algebra we arrive at the two body light-cone
wavefunctions of the proton with
\begin{equation}
\Psi_F^{\uparrow,\downarrow}=\frac{1}{\sqrt{2}}|u\,S\rangle_F^{\Uparrow,\Downarrow}
+\frac{1}{\sqrt{6}} |u\,
V\rangle_F^{\Uparrow,\Downarrow}-\frac{1}{\sqrt{3}}|d\,V\rangle_F^{\Uparrow,\Downarrow}.\label{eq:lcwf}
\end{equation}

\begin{figure}[b]
\begin{center}
\scalebox{0.35}{\includegraphics*[3pt,46pt][592pt,508pt]{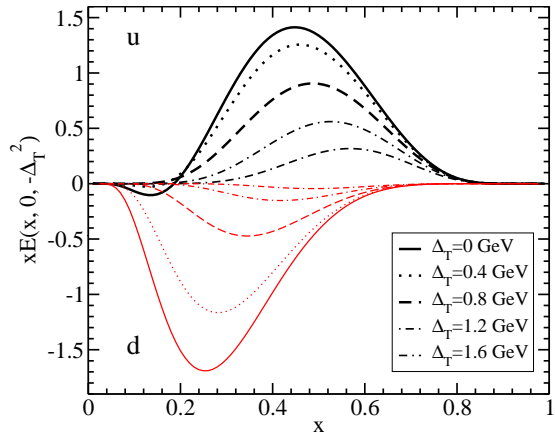}}
\caption{\small The generalized parton distributions
$E_u(x,0,\Delta_T^2)$ and $E_d(x,0,\Delta_T^2)$ for the
proton in the light-cone diquark model as functions of x for
different values of $\Delta_T$.}\label{E-ud}
\end{center}
\end{figure}

The scalar diquark component of the wavefunction for the proton has
the form
\begin{eqnarray}
|u\,S(P^+,\kT )\rangle^{\Uparrow,\Downarrow}&=&\sum_{s_z=\pm
\frac{1}{2}} \int \frac{d^2\kT
d x}{\sqrt{x(1-x)}16\pi^3}
\nonumber \\
&
 \times&\psi^{\Uparrow,\Downarrow}_S(x,\kT ,s_z)|xP^+,\kT
,s_z\rangle,\label{eq:sswf}
\end{eqnarray}
while the vector diquark component is expressed as
\begin{eqnarray}
|q\, V(P^+,\kT )\rangle^{\Uparrow,\Downarrow}&=&\sum_{l_z=0,\pm
1;\,s_z=\pm \frac{1}{2}} \int\frac{d^2\kT
d x}{\sqrt{x(1-x)}16\pi^3} \nonumber
\\
&
 \times&\psi^{\Uparrow,\Downarrow}_V(x,\kT ,l_z,s_z)|xP^+, \kT
,l_z,s_z\rangle ,\label{eq:svwf}
\end{eqnarray}
which is the same for $|u\,V\rangle_F$ and $|d\,V\rangle_F$.  Here
we denote $s_z$ and $l_z$ as the spin projections of the quark and
the vector diquark. The forms of $\psi^{\Uparrow,\Downarrow}_S(x,\kT
,s_z)$ and $\psi^{\Uparrow,\Downarrow}_V(x,\kT ,l_z,s_z)$ are given in
the appendix.

Now we calculate the chiral-even GPDs in the zero skewedness
($\xi=0$) where $t=-\boldsymbol{\Delta_T}^2$. In the overlap
representation~\cite{Diehl:2000xz,Brodsky:2000xy} $H$, $E$ and
$\widetilde{H}$ at $\xi=0$ can be expressed in a symmetric frame as ( in
the domain $0<x<1$ and for $n\rightarrow n$ transition):
\begin{widetext}
\begin{eqnarray}
H(x,0,-\boldsymbol{\Delta}_T^2) &=& \sum_{n,\lambda_i}\int
\prod_{i=1}^n\frac{dx_id^2\boldsymbol{k}_{Ti}}{16\pi^3}
16\pi^3\delta\left(1-\sum_{j=1}^n x_j\right)
\delta^{(2)}\left(\sum_{j=1}^{n}\boldsymbol{k}_{Tj}\right)\delta(x-x_1)\psi_n^{\uparrow
\star}(x_i^\prime, \boldsymbol{k}_{Ti}^\prime, \lambda_i)
\psi_n^{\uparrow }(y_i, \boldsymbol{l}_{Ti}, \lambda_i),\\
\frac{\boldsymbol{\Delta}_L}{2M}E(x,0,-\boldsymbol{\Delta}_T^2)&=&
\sum_{n,\lambda_i}\int
\prod_{i=1}^n\frac{dx_id^2\boldsymbol{k}_{Ti}}{16\pi^3}
16\pi^3\delta\left(1-\sum_{j=1}^n x_j\right)
\delta^{(2)}\left(\sum_{j=1}^{n}\boldsymbol{k}_{Tj}\right)\delta(x-x_1)\psi_n^{\uparrow
\star}(x_i^\prime, \boldsymbol{k}_{Ti}^\prime, \lambda_i)
\psi_n^{\downarrow }(y_i, \boldsymbol{l}_{Ti}, \lambda_i),\label{e-zero-skew}\\
\widetilde{H}(x,0,-\boldsymbol{\Delta}_T^2) &=& \sum_{n,\lambda_i}\int
\prod_{i=1}^n
\textrm{sign}(\lambda_i)\frac{dx_id^2\boldsymbol{k}_{Ti}}{16\pi^3}
16\pi^3\delta\left(1-\sum_{j=1}^n x_j\right)
\delta^{(2)}\left(\sum_{j=1}^{n}\boldsymbol{k}_{Tj}\right)\delta(x-x_1)\psi_n^{\uparrow
\star}(x_i^\prime, \boldsymbol{k}_{Ti}^\prime, \lambda_i)
\psi_n^{\uparrow }(y_i, \boldsymbol{l}_{Ti}, \lambda_i),\nonumber\\
\end{eqnarray}
\end{widetext}
with
\begin{eqnarray}
x^\prime_1 &=&
x_1,~~~~~~\boldsymbol{k}_{T1}^{\,\prime}=\boldsymbol{k}_{T1}-(1-x_1)\frac{\boldsymbol{\Delta_T}}{2}
\nonumber \\
&&~~~~~\textrm{for the final struck quark,}\nonumber\\
x^\prime_i &=&
x_i,~~~~~~~\boldsymbol{k}_{Ti}^{\,\prime}=\boldsymbol{k}_{Ti}+x_i\frac{\boldsymbol{\Delta_T}}{2}\nonumber \\
&&~~~~~\textrm{for the final (n-1) spectators,}\nonumber
\end{eqnarray}
and
\begin{eqnarray}
y_1 &=&
x_1,~~~~~~\boldsymbol{l}_{T1}^{\,\prime}=\boldsymbol{k}_{T1}+(1-x_1)\frac{\boldsymbol{\Delta_T}}{2}
\nonumber \\
&&~~~~\textrm{for the initial struck quark,}\nonumber\\
y_i &=&
x_i,~~~~~~~\boldsymbol{l}_{Ti}=\boldsymbol{k}_{Ti}-x_i\frac{\boldsymbol{\Delta_T}}{2}
\nonumber \\
&&~~~~~\textrm{for the initial (n-1) spectators,}\nonumber
\end{eqnarray}

From Eq.~(\ref{e-zero-skew}) we see that non-zero $E_q$ needs a spin
flip between the initial and final proton wavefunctions. The same
kind of overlap integration of light-front wavefunctions (with
$J_z=\pm 1/2$ in the initial and final states) also appears in the
calculation~\cite{ref:bhs} of Sivers functions, which indicates the
presence of the quark OAM .

By employing the light-cone wavefunctions given in (\ref{eq:lcwf})
and the overlap representation formalism, we calculate the
generalized parton distribution functions (GPDs)
$H_q(x,0,-\boldsymbol{\Delta}_T^2)$,
$\widetilde{H}_q(x,0,-\boldsymbol{\Delta}_T^2)$ and
$E_q(x,0,-\boldsymbol{\Delta}_T^2)$ at zero skewedness for $q=u$
and $d$ quarks. The $x$-dependence of these GPDs at different values
of $\Delta_T$ are given in Figs.~\ref{xH-ud}, \ref{Htidle-ud} and
\ref{E-ud}, respectively.

From Fig.~\ref{E-ud} one can see that $E_u$ and $E_d$ have opposite
sign ($E_u$ is positive and $E_d$ is negative) with similar size in
our model. Since it has been shown that there is a quantitative
relation~\cite{amm,ls07,gpdtmd} between the Sivers function
$f_{1T}^{\bot q}$ and the GPD $E^q$, our result coincides
with recent
extractions~\cite{Anselmino:2005ea,Efremov:2004tp,Vogelsang:2005cs}
of the Sivers function from the Semi-inclusive deeply inelastic scattering
data, which show the Sivers
functions of $u$ and $d$ have opposite sign with similar size.

Special attention should be paid to the limit of zero momentum
transfer $\Delta_T^2=0$, since in this limit the GPDs $H_q$ and
$\widetilde{H}_q$ are simplified to the forward distribution $q(x)$ and
$\Delta_q(x)$. Also the quark OAM s are related in the way shown in
(\ref{lqz}), from which in principle one can calculates $L_q(x)$
from the known chiral-even GPDs.

\begin{figure}
\begin{center}
\scalebox{0.3}{\includegraphics*[10pt,30pt][593pt,508pt]{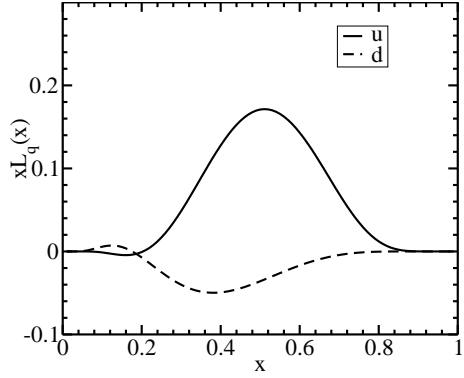}}
\caption{\small The OAM distributions $L_q(x)$
of $u$ and $d$ quarks inside the proton in the light-cone
diquark model as functions of $x$.}\label{xL-ud}
\end{center}
\end{figure}

By taking the GPDs in the forward limit, we calculate the
OAM distributions of u and d quarks inside the proton,
as shown in Fig.~\ref{xL-ud}. It can be seen that in our model
$L_u(x)$ is positive,  while $L_d(x)$ is consistent with zero
compared with $L_u(x)$, and the net OAM of the
$u$ and $d$ quarks is positive. From Fig.~\ref{E-ud} one can see
that $E_d$ is sizable. However, since $E_d$ is negative, there is a
cancelation between $d(x)$, $E_d(x)$ and $\Delta q(x)$. This leads
to a small contribution of the $d$ quark orbital angular momentum.
We remind that there are Lattice
QCD~\cite{Hagler:2003jd,Gockeler:2003jfa}, as well as
phenomenological parametrizations and other model calculations of
GPDs
~\cite{Ossmann:2004bp,Guidal:2004nd,Diehl:2004cx,Wakamatsu:2005vk,Wakamatsu:2006dy},
which are used to estimate the OAM of the quarks.


\begin{figure}
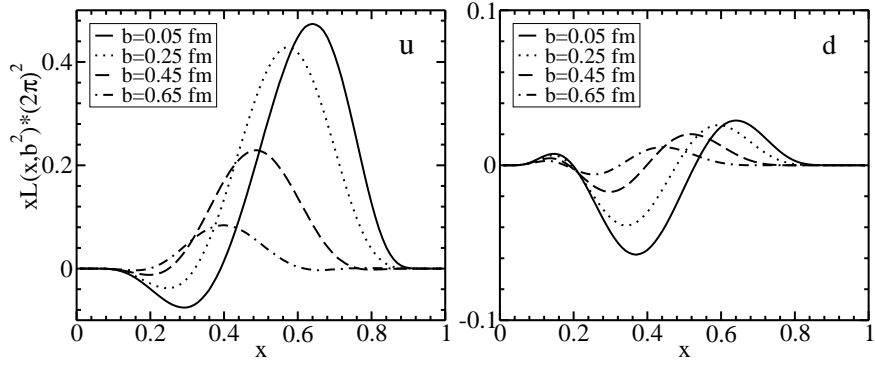

\begin{center}
\scalebox{0.28}{\includegraphics*[0pt,38pt][596pt,509pt]{fig/xL-u-x-difb.eps}}
\scalebox{0.28}{\includegraphics*[33pt,38pt][594pt,519pt]{fig/xL-d-x-difb.eps}}
\caption{\small The impact parameter distributions (scaled with a factor of $(2\pi)^2$) $xL_u(x,\bT)$
(left) and $xL_d(x,\bT)$ (right) for the proton in the light-cone
diquark model as functions of $x$ for different values of
$b$.}\label{xL-ud-x-difb}
\end{center}
\end{figure}

\begin{figure*}
\begin{center}
\scalebox{0.34}{\includegraphics*[0pt,0pt][586pt,520pt]{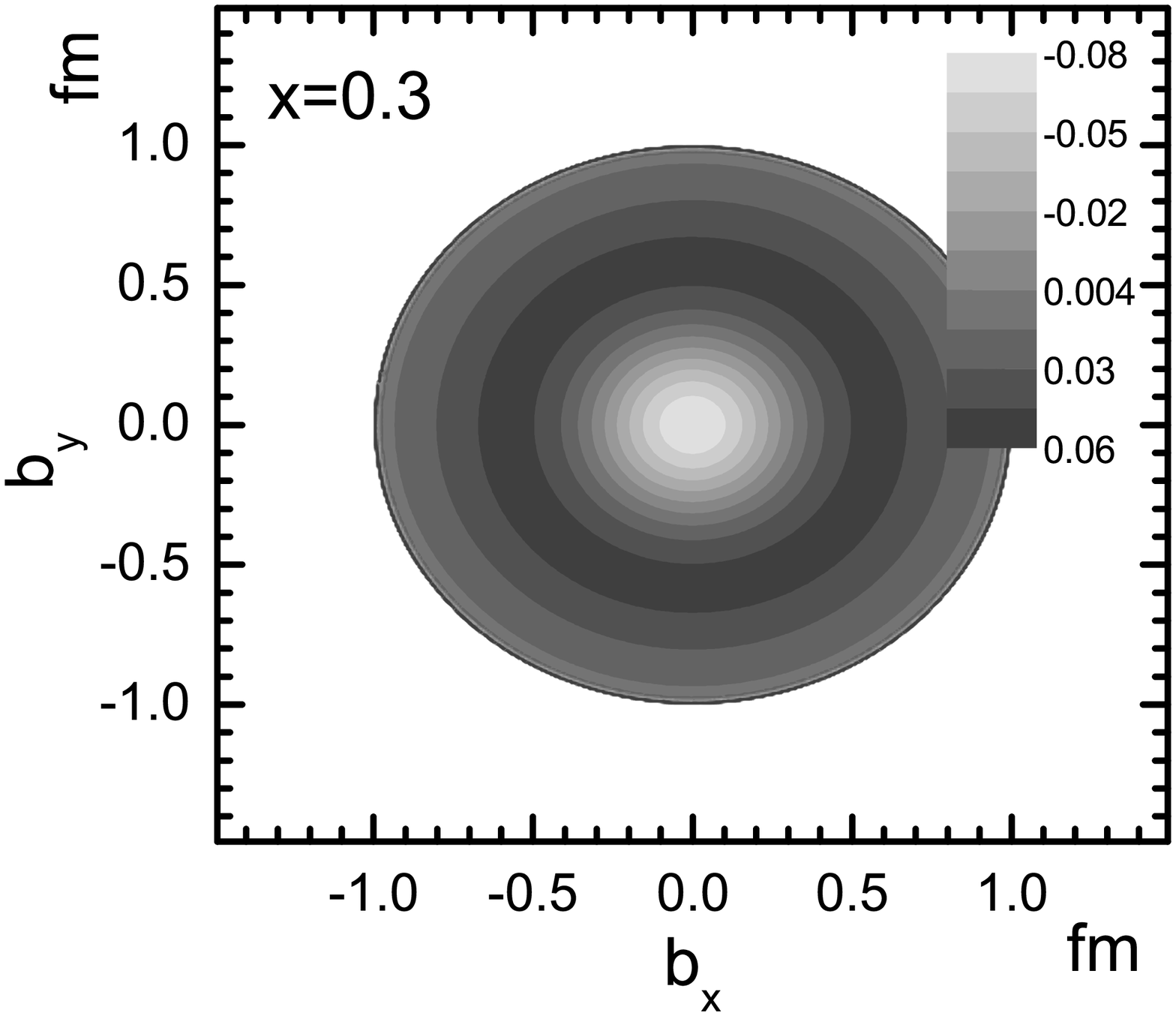}}
\scalebox{0.34}{\includegraphics*[0pt,0pt][586pt,520pt]{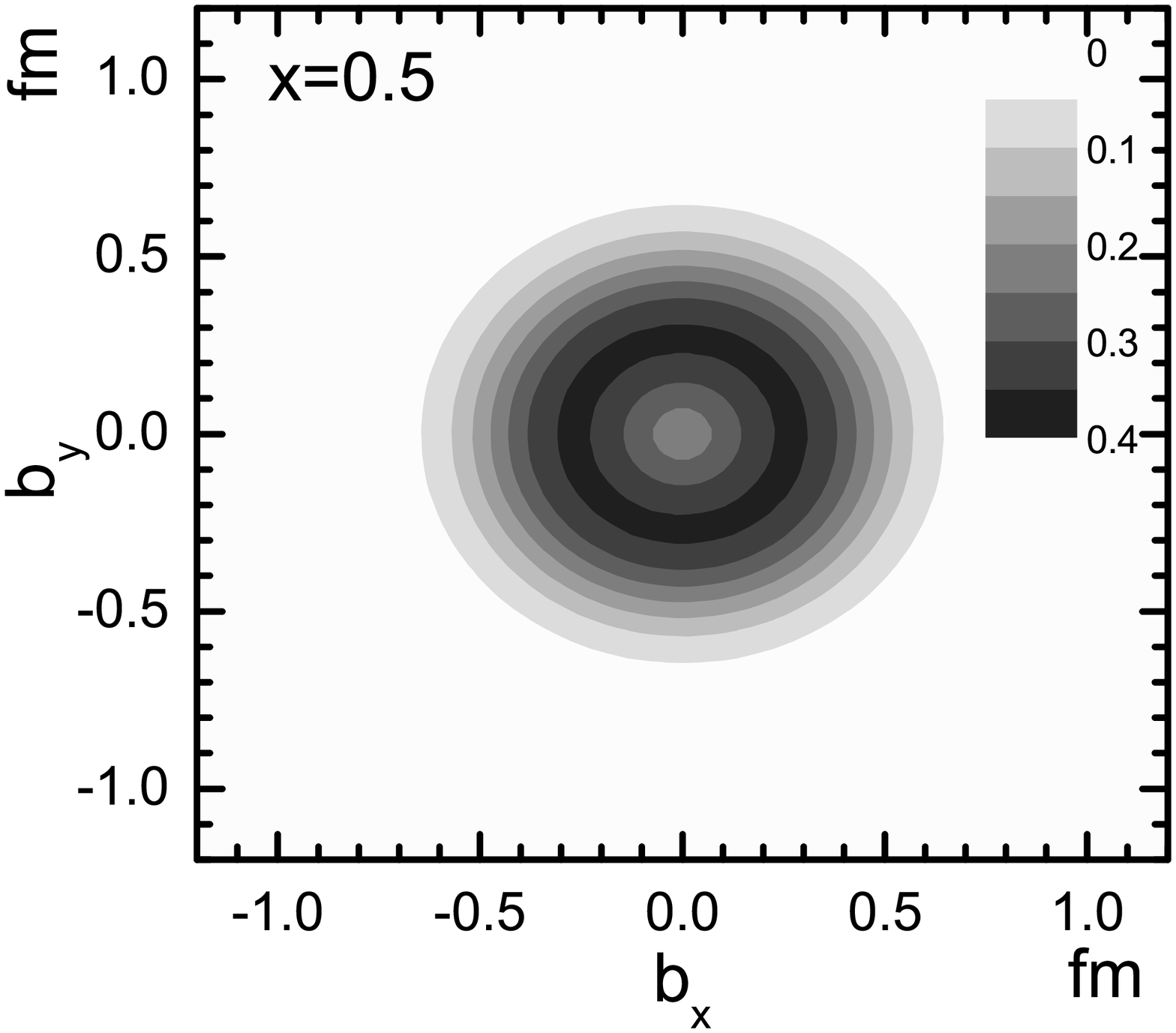}}
\caption{\small The profiles of the impact parameter distribution (scaled by a factor of $(2\pi)^2$)
$xL_u(x,\bT)$ for the proton in the light-cone diquark model as
functions of $\Delta_T$ for $x=0.3$ (left) and $x=0.5$
(right).}\label{xL-u-impact-b}
\end{center}
\end{figure*}

\section{Impact parameter dependence of Orbital angular momentum}

In this section we want to study the quark OAM s in transverse
position (impact parameter) space. The GPDs in the impact parameter
space have been studied in
Refs.~\cite{Burkardt:2000za,Diehl:2002he,Burkardt:2002hr}. The most
interesting case is the zero skewedness limit $\xi = 0$, in which a
density interpretation of GPDs in the impact parameter space may be
obtained~\cite{Burkardt:2000za}, Therefore studying GPDs in
impact parameter space can provide a three-dimensional picture of
the nucleon. In the following we restrict ourselves to the case $\xi=0$.

The impact parameter PDFs inside the nucleon can be obtained by
sandwiching the parton correlator between nucleon states
localized in  transverse space
\begin{eqnarray}
q(x,\bT) &=&\;\big<P^+,\nT;S\big|
\hat{\mathcal{O}}^{[\gamma^+]}_q(x,\bT)
 \big|P^+,\nT;S\big>,
\end{eqnarray}
where
\begin{eqnarray}
&&\hat{{O}}^{[\gamma^+]}_q(x,\bT)\nonumber\\
&=&\int \frac{d y^-}{8\pi}\;e^{ix P^+y^-/2}\;
\bar\psi(0,-\frac{y^-}{2},\bT)\,\gamma^+\,\psi(0,\frac{y^-}{2},\bT),
\end{eqnarray}
and the initial and final states in the transverse space defined as
~\cite{Soper:1976jc,Burkardt:2000za,Diehl:2002he}
\begin{align} \label{e:ini}
\big|P^+,\bT;S\big>&=\mathcal{N}\int\frac{d^2\pT}{(2\pi)^2}\,e^{-i\pT\cdot\bT}\,\big|p;S\big>
\,,
\\ \label{e:out}
\big<P^+,\bT;S\big|&=\mathcal{N^*}\int\frac{d^2\ppT}{(2\pi)^2}\,e^{i\ppT\cdot\bT}\,\big<p';S\big|
\,,
\end{align}
which characterize a nucleon with momentum $P^+$ at a transverse
position $\bT$ and polarization specified by $S$.

One of the interesting features of impact parameter dependent parton
distributions is that they are Fourier transformations of
GPDs~\cite{Burkardt:2000za}. For instance, The impact parameter
dependence of unpolarized quark in the unpolarized nucleon can be
obtained from
\begin{align}
q(x,\bT)=\int \frac{d^2\DT}{(2\pi)^2} e^{-i\bT \cdot \DT}
H_q(x,0,-\DT^2),
\end{align}
here $\bT$ and $\DT$ are two conjugated parameters.

Similarly the impact parameter dependence of quark helicity
distribution in the longitudinal polarized nucleon is defined as
\begin{eqnarray}
\Delta q(x,\bT) &=&\;\big<P^+,\nT;S\big|
\hat{\mathcal{O}}^{[\gamma^+\gamma_5]}_q(x,\bT)
 \big|P^+,\nT;S\big>,
\end{eqnarray}
Which is the Fourier transformation of $\widetilde{H}_q$:
\begin{align}
q(x,\bT)=\int \frac{d^2\DT}{(2\pi)^2} e^{-i\bT \cdot \DT}
H_q(x,0,-\DT^2),
\end{align}

We follow a similar approach by introducing the impact parameter
dependence of quark OAM  $\mathcal{L}(x,\bT)$. It can be obtained
from the expectation value of $\hat{O}_{L}$, given in
Eq.~(\ref{oam-operator}), between the position state $|P^+,
\boldsymbol{0}_T \rangle $:
\begin{eqnarray}
L_q(x,\bT)= \left\langle P,\nT; S \left | \hat{O}_{L} \right
|P,\nT;S  \right\rangle
\end{eqnarray}
After a Fourier transformation on $L_q(x,\bT)$ one can arrive at
\begin{equation}
\int d^2\boldsymbol b_T \, \textrm{e}^{\textrm{i}
\boldsymbol{b}_T\,\cdot \,\boldsymbol\Delta_T}\,L_q^z(x,\boldsymbol
b_T) = L_q(x,-\boldsymbol \Delta^2_T)
\end{equation}
The function $L_q^z(x,-\boldsymbol \Delta^2_T)$ can be obtained by
the GPDs at zero skewedness~\cite{ji97}
\begin{eqnarray}
L_q(x,-\boldsymbol \Delta^2_T)&=&
\frac{1}{2}\left\{x\left[H(x,0,-\boldsymbol
\Delta^2_T)+E_q(x,0,-\boldsymbol
\Delta^2_T)\right]\right. \nonumber\\
&-&\left.\widetilde{H}(x,0,-\boldsymbol \Delta^2_T)\right\},
\end{eqnarray}
and (\ref{lqz}) is the forward limit of $L_q(x,-\boldsymbol
\Delta^2_T)$.

Therefore, if we know the GPDs $H_q$ $\widetilde{H}_q$ and $E_q$, from
(\ref{lqz}) one can calculate the impact parameter dependence of the quark
OAM distribution by the Fourier transformation
\begin{equation}
L_q(x,\boldsymbol b_T) = \int {d^2\DT \over(2\pi)^2}\,
\textrm{e}^{-\textrm{i} \boldsymbol{b}_T\,\cdot
\,\boldsymbol\Delta_T}\, L_q^z(x,\boldsymbol \Delta^2_T).
\end{equation}
The integration over impact parameter dependence of quark OAM  leads
to
\begin{equation}
\int d^2\boldsymbol{b}_T L_q(x,\boldsymbol{b}_T) =L_q(x)
\end{equation}

In Fig.~\ref{xL-ud-x-difb} we shown the impact parameter
distributions (scaled with a factor of $(2\pi)^2$) $L_u(x,\bT)$ (left) and $L_d(x,\bT)$ (right) for the
proton in the light-cone diquark model, as functions of $x$, for
different values of $b$. In Fig.~\ref{xL-u-impact-b} we show the
profiles of the impact parameter distributions $L_u(x,\bT)$ for the
proton in the light-cone diquark model as functions of $\bT$,
for $x=0.3$ and $x=0.5$. It is shown that the impact parameter
dependence of quark OAM  is axially symmetric. Also at large $x$ the
impact parameter distribution is peaked at small $b$.

\section{summary}

As a conclusion, we study the OAM structure of
the quarks inside the proton in a light-cone diquark model. In this
model the light-cone wave function of the proton is known. It is
then convenient to express the physical observables in the overlap
representation formalism. We calculate the chiral-even generalized
parton distribution functions (GPDs) $H_q(x,\xi,\Delta^2)$,
$\widetilde{H}_q(x,\xi,\Delta^2)$ and $E_q(x,\xi,\Delta^2)$ at zero
skewedness for $q=u$ and $d$. We found that $E_u$ and $E_d$ have
opposite sign, with similar size in our model. The GPDs are applied
to calculate the OAM distributions, showing
that $L_u(x)$ is positive, while $L_d(x)$ is consistent with zero
compared with $L_u(x)$, and the net OAM of the
$u$ and $d$ quarks is positive. We also introduce the impact
parameter dependence of quark OAM  distribution  $L(x,\bT)$ . It
describes the position space distribution of the quark OAM at given $x$.
We found that the impact parameter
dependence of quark OAM distribution is axially
symmetric in the light-cone diquark model.

\section*{ACKNOWLEDGMENTS}
This work is supported by FONDECYT (Chile) Project No. 11090085 and
No.1100715.

\appendix

\section{light-cone wave functions in a diquark model}

The expressions for $\psi^{\Uparrow,\Downarrow}_S(x,\kT
,s_z)$ have the form
\begin{eqnarray}
\psi^\Uparrow_S(x,\kT ,+\frac{1}{2})&=&\frac{(k^++m)}{\omega}\phi_S(x,k_T ),\nonumber\\
\psi^\Uparrow_S(x,\kT ,-\frac{1}{2})&=&-\frac{k_r}{\omega}
 \phi_S(x,k_T ),\label{eq:cfs}
\end{eqnarray}
and
\begin{eqnarray}
\psi^\Downarrow_S(x,\kT ,+\frac{1}{2})&=&\frac{k_l}{\omega}\phi_S(x,k_T ),\nonumber\\
\psi^\Downarrow_S(x,\kT ,-\frac{1}{2})&=&\frac{(k^++m)}
{\omega}\phi_S(x,k_T ),\label{eq:cfs}
\end{eqnarray}
respectively.

The expressions of $\psi^{\Uparrow,\Downarrow}_V(x,\kT
,l_z,s_z)$ can be expressed as
\begin{eqnarray}
\psi^\Uparrow_V(x,\kT ,+1,\uparrow)&=&-\sqrt{2}
\frac{\phi_V(x,k_T )}{\omega \omega^2_V}\bigg{[}(k_V^++\lambda_V)(k^++m)\nonumber\\
&&+(k_V^++\lambda_V)^2\bigg{]}k^L,\nonumber\\
\psi^\Uparrow_V(x,\kT ,+1,\downarrow)&=&
\sqrt{2}\frac{\phi_V(x,k_T )}{\omega
\omega^2_V}\bigg{[}(k_V^++\lambda_V)\kT^2
\nonumber\\
&&-(k_V^++\lambda_V)^2(k^++m)\bigg{]},\nonumber\\
\psi^\Uparrow_V(x,\kT ,0,\uparrow)&=& 2\frac{\phi_V(x,k_T
)}{\omega
\omega^2_V}\left\{\bigg{[}(k_V^0+\lambda_V)k_V^+-\kT^2
\bigg{]}(k^++m)\right.
\nonumber\\
&&\left.-(k_V^++\lambda_V)\kT^2 \right\},\label{eq:vec}\\
\psi^\Uparrow_V(x,\kT ,0,\downarrow)&=& \frac{\phi_V(x,k_T
)}{\omega
\omega^2_V}\bigg{[}-2((k_V^0+\lambda_V)k_V^+-\kT^2 )
\nonumber\\
&&-2(k_V^++\lambda_V)(k^++m)\bigg{]}k^R,\nonumber\\
\psi^\Uparrow_V(x,\kT ,-1,\uparrow)&=&\sqrt{2}
\frac{\phi_V(x,k_T )}{\omega \omega^2_V}\bigg{[}(k_V^++\lambda_V)(k^++m)- \kT^2 \bigg{]}k^R,\nonumber\\
\psi^\Uparrow_V(x,\kT ,-1,\downarrow)&=&-\sqrt{2}
\frac{\phi_V(x,k_T )}{\omega \omega^2_V}\bigg{[}k^{R2}_T
(k_V^++\lambda_V+k^++m)\bigg{]},\nonumber
\end{eqnarray}
and
\begin{eqnarray}
\psi^\Downarrow_V(x,\kT ,+1,\uparrow)&=& -\sqrt{2}
\frac{\phi_V(x,k_T )}{\omega \omega^2_V}\bigg{[}k^{L2}_T
(k_V^++\lambda_V+k^++m)\bigg{]},
\nonumber\\
\psi^\Downarrow_V(x,\kT
,+1,\downarrow)&=&-\sqrt{2} \frac{\phi_V(x,k_T )}{\omega
\omega^2_V}\bigg{[}(k_V^++\lambda_V)(k^++m)\nonumber \\
&&-
\kT^2 \bigg{]}k^L,\nonumber\\
\psi^\Downarrow_V(x,\kT ,0,\uparrow)&=&2\frac{\phi_V(x,k_T
)}{\omega \omega^2_V}\bigg{[}((k_V^0+\lambda_V)k_V^+-\kT^2
)\nonumber\\
&&+(k_V^++\lambda_V)(k^++m)\bigg{]}k^L,\\
\psi^\Downarrow_V(x,\kT ,0,\downarrow)&=&2
\frac{\phi_V(x,k_T )}{\omega
\omega^2_V}\left\{\bigg{[}(k_V^0+\lambda_V)k_V^+
-\kT^2 \bigg{]}(k^++m)\nonumber\right.\\
&&-\left.(k_V^++\lambda_V)\kT^2 \right\},\nonumber\\
\psi^\Downarrow_V(x,\kT
,-1,\uparrow)&=&\sqrt{2}\frac{\phi_V(x,k_T )}{\omega
\omega^2_V}\bigg{[}(k_V^++\lambda_V)\kT^2 \nonumber \\
&&-
(k_V^++\lambda_V)^2(k^++m)\bigg{]},\nonumber\\
\psi^\Downarrow_V(x,\kT
,-1,\downarrow)&=&\sqrt{2}\frac{\phi_V(x,k_T )}{\omega
\omega^2_V}\bigg{[}(k_V^++\lambda_V)(k^++m)\nonumber \\
&&+ (k_V^++\lambda_V)^2\bigg{]}k^R,\nonumber
\end{eqnarray}

The momentum dependence of the wavefunctions in the above equations
is described by $\phi_D(x,k_T^2)$ with the Gaussian form
\begin{eqnarray}
\phi_D(x,k_T)=A_D\exp\left
(-\frac{\mathcal{M}^2}{8\beta_D^2}\right ),
\end{eqnarray}
where
\begin{equation}
\mathcal{M}^2=\frac{\kT^2+m_q^2}{x}+\frac{\kT^2+\lambda_V^2}{1-x},
\end{equation}
  $A_D$ stands for the normalization constant, and $\beta_D$ is the oscillation factor. For the
parameters we adopt the values from \cite{Ma:2002ir}, which can describe the data of the nucleon form factors.


\begin{thebibliography}{99}

\bibitem{emc} EMC Collaboration, J. Ashman et al.,Phys.Lett. B202 (1988) 603;
Nucl. Phys. B 328 (1989) 1.

\bibitem{Sehgal:1974rz}
  L.~M.~Sehgal,
  Phys.\ Rev.\  D {\bf 10}, 1663 (1974)
  [Erratum-ibid.\  D {\bf 11}, 2016 (1975)].

\bibitem{orbital} R. L. Jaffe and A. Manohar, Nucl. Phys. B 337 (1990)
509.
\bibitem{ji97} X. Ji, Phys. Rev. Lett. 78 (1997) 610.
\bibitem{hagler98} P. Hagler and A. Schafer, Phys. Lett. B 430 (1998) 179.
\bibitem{kundu99} A. Harindranath and R. Kundu, Phys. Rev. D 59 (1999) 116013.




\bibitem{Ma:1998ar}
  B.~Q.~Ma and I.~Schmidt,
  Phys.\ Rev.\  D {\bf 58}, 096008 (1998)
  [arXiv:hep-ph/9808202].

\bibitem{Mueller:1998fv}
  D.~Mueller, D.~Robaschik, B.~Geyer, F.~M.~Dittes and J.~Horejsi,
  Fortsch.\ Phys.\  {\bf 42}, 101 (1994)
  [arXiv:hep-ph/9812448].

\bibitem{Ji:1996nm}
  X.~D.~Ji,
  Phys.\ Rev.\  D {\bf 55}, 7114 (1997)
  [arXiv:hep-ph/9609381].

\bibitem{Radyushkin:1997ki}
  A.~V.~Radyushkin,
  Phys.\ Rev.\  D {\bf 56}, 5524 (1997)
  [arXiv:hep-ph/9704207].

\bibitem{diehl03} M. Diehl, Phys. Rept. 388 (2003) 41.

\bibitem{Belitsky:2005qn}
  A.~V.~Belitsky and A.~V.~Radyushkin,
  Phys.\ Rept.\  {\bf 418}, 1 (2005)
  [arXiv:hep-ph/0504030].

\bibitem{Boffi:2007yc}
  S.~Boffi and B.~Pasquini,
  Riv.\ Nuovo Cim.\  {\bf 30}, 387 (2007)
  [arXiv:0711.2625 [hep-ph]].

\bibitem{Radyushkin:1996nd}
  A.~V.~Radyushkin,
  Phys.\ Lett.\  B {\bf 380}, 417 (1996)
  [arXiv:hep-ph/9604317].

\bibitem{Polyakov:1998ze}
  M.~V.~Polyakov,
  Nucl.\ Phys.\  B {\bf 555}, 231 (1999)
  [arXiv:hep-ph/9809483].

\bibitem{Collins:1996fb}
  J.~C.~Collins, L.~Frankfurt and M.~Strikman,
  Phys.\ Rev.\  D {\bf 56}, 2982 (1997)
  [arXiv:hep-ph/9611433].

\bibitem{Mulders:1995dh}
  P.~J.~Mulders and R.~D.~Tangerman,
  Nucl.\ Phys.\  B {\bf 461}, 197 (1996)
  [Erratum-ibid.\  B {\bf 484}, 538 (1997)]
  [arXiv:hep-ph/9510301].
\bibitem{Boer:1997nt}
  D.~Boer and P.~J.~Mulders,
  Phys.\ Rev.\  D {\bf 57}, 5780 (1998)
  [arXiv:hep-ph/9711485].

\bibitem{Sivers:1989cc}
  D.~W.~Sivers,
  Phys.\ Rev.\  D {\bf 41}, 83 (1990).

\bibitem{Sivers:1990fh}
  D.~W.~Sivers,
  Phys.\ Rev.\  D {\bf 43}, 261 (1991).


\bibitem{Airapetian:2004tw}

A.~Airapetian {\it et al}. (HERMES Collaboration), Phys. Rev. Lett.
{\bf 94}, 012002 (2005).

\bibitem{hermes05} 
  M.~Diefenthaler  [HERMES Collaboration],
  AIP Conf.\ Proc.\  {\bf 792}, 933 (2005)
  [arXiv:hep-ex/0507013].



\bibitem{compass}
V.Yu.~Alexakhin {\it et al}. (COMPASS Collaboration), Phys. Rev.
Lett. {\bf 94}, 202002 (2005).

\bibitem{compass06}  E.~S.~Ageev {\it et al.}  [COMPASS Collaboration],
  Nucl.\ Phys.\  B {\bf 765}, 31 (2007)
  [arXiv:hep-ex/0610068].


\bibitem{amm}  M. Burkardt and D. S. Hwang, Phys. Rev. D 69, 074032
(2004);
M. Burkardt, ibid. D 72, 094020 (2005);

\bibitem{Burkardt:2005km}
  M.~Burkardt and G.~Schnell,
  Phys.\ Rev.\  D {\bf 74}, 013002 (2006)
  [arXiv:hep-ph/0510249].
\bibitem{ls07}
  Z.~Lu and I.~Schmidt,
  Phys.\ Rev.\  D {\bf 75}, 073008 (2007)
  [arXiv:hep-ph/0611158].

\bibitem{Diehl:2005jf}
  M.~Diehl and Ph.~Hagler,
  Eur.\ Phys.\ J.\  C {\bf 44}, 87 (2005)
  [arXiv:hep-ph/0504175].

\bibitem{Gockeler:2006zu}
  M.~Gockeler {\it et al.}  [QCDSF Collaboration and UKQCD Collaboration],
  Phys.\ Rev.\ Lett.\  {\bf 98}, 222001 (2007)
  [arXiv:hep-lat/0612032].

\bibitem{gpdtmd} S. Meissner, A. Metz, and K. Goeke, Phys. Rev. D 76, 034002 (2007).


\bibitem{Meissner:2008ay}
  S.~Meissner, A.~Metz, M.~Schlegel and K.~Goeke,
  JHEP {\bf 0808}, 038 (2008)
  [arXiv:0805.3165 [hep-ph]].

\bibitem{Meissner:2009ww}
  S.~Meissner, A.~Metz and M.~Schlegel,
  JHEP {\bf 0908}, 056 (2009)
  [arXiv:0906.5323 [hep-ph]].

\bibitem{Burkardt:2002ks}
  M.~Burkardt,
  Phys.\ Rev.\  D {\bf 66}, 114005 (2002)
  [arXiv:hep-ph/0209179].

\bibitem{Burkardt:2003uw}
  M.~Burkardt,
  Nucl.\ Phys.\  A {\bf 735}, 185 (2004)
  [arXiv:hep-ph/0302144].






\bibitem{Burkardt:2000za}
  M.~Burkardt,
  Phys.\ Rev.\  D {\bf 62}, 071503 (2000)
  [Erratum-ibid.\  D {\bf 66}, 119903 (2002)]
  [arXiv:hep-ph/0005108].

\bibitem{Burkardt:2002hr}
  M.~Burkardt,
  Int.\ J.\ Mod.\ Phys.\  A {\bf 18}, 173 (2003)
  [arXiv:hep-ph/0207047].

\bibitem{Diehl:2002he}
  M.~Diehl,
  Eur.\ Phys.\ J.\  C {\bf 25}, 223 (2002)
  [Erratum-ibid.\  C {\bf 31}, 277 (2003)]
  [arXiv:hep-ph/0205208].


\bibitem{Burkardt:2003je}
  M.~Burkardt and D.~S.~Hwang,
  Phys.\ Rev.\  D {\bf 69}, 074032 (2004)
  [arXiv:hep-ph/0309072].

\bibitem{Brodsky:2000xy}
  S.~J.~Brodsky, M.~Diehl and D.~S.~Hwang,
  Nucl.\ Phys.\  B {\bf 596}, 99 (2001)
  [arXiv:hep-ph/0009254].

\bibitem{Diehl:2000xz}
  M.~Diehl, T.~Feldmann, R.~Jakob and P.~Kroll,
  Nucl.\ Phys.\  B {\bf 596}, 33 (2001)
  [Erratum-ibid.\  B {\bf 605}, 647 (2001)]
  [arXiv:hep-ph/0009255].

\bibitem{Burkardt:2008ua}
  M.~Burkardt and B.~C.~Hikmat,
  Phys.\ Rev.\  D {\bf 79}, 071501 (2009)
  [arXiv:0812.1605 [hep-ph]].






\bibitem{Hoodbhoy:1998yb}
  P.~Hoodbhoy, X.~D.~Ji and W.~Lu,
  Phys.\ Rev.\  D {\bf 59}, 014013 (1999)
  [arXiv:hep-ph/9804337].

\bibitem{ref:qdq} M.I. Pavkovi$\acute{\textmd{c}}$, Phys. Rev. D 13
(1976) 2128.
\bibitem{Ma}
  B.~Q.~Ma,
  Phys.\ Lett.\  B {\bf 375}, 320 (1996)
  [Erratum-ibid.\  B {\bf 380}, 494 (1996)]
  [arXiv:hep-ph/9604423].



\bibitem{Ma:2002ir}
  B.~Q.~Ma, D.~Qing and I.~Schmidt,
  Phys.\ Rev.\  C {\bf 65}, 035205 (2002)
  [arXiv:hep-ph/0202015].


\bibitem{ref:melosh} H.J. Melosh, Phys. Rev. D 9
(1974) 1095.

\bibitem{ref:as} D.V. Ahluwalia and M. Sawicki, Phys. Rev. D 47
(1993) 5161.


\bibitem{weinberg64} S. Weinberg, Phys. Rev. B133, 1318 (1964); D. E. Soper, Ph.D.
thesis, SLAC (1971).
\bibitem{ref:bhs} S.J. Brodsky, D.S. Hwang, and I. Schmidt, Phys. Lett. B 530 (2002) 99.

\bibitem{Anselmino:2005ea}
  M.~Anselmino, M.~Boglione, U.~D'Alesio, A.~Kotzinian, F.~Murgia and A.~Prokudin,
  Phys.\ Rev.\  D {\bf 72}, 094007 (2005)
  [Erratum-ibid.\  D {\bf 72}, 099903 (2005)]
  [arXiv:hep-ph/0507181].

\bibitem{Efremov:2004tp}
  A.~V.~Efremov, K.~Goeke, S.~Menzel, A.~Metz and P.~Schweitzer,
  Phys.\ Lett.\  B {\bf 612}, 233 (2005)
  [arXiv:hep-ph/0412353].

\bibitem{Vogelsang:2005cs}
  W.~Vogelsang and F.~Yuan,
  Phys.\ Rev.\  D {\bf 72}, 054028 (2005)
  [arXiv:hep-ph/0507266].


\bibitem{Hagler:2003jd}
  P.~Hagler, J.~W.~Negele, D.~B.~Renner, W.~Schroers, T.~Lippert and K.~Schilling
                  [LHPC collaboration and SESAM collaboration],
  Phys.\ Rev.\  D {\bf 68}, 034505 (2003)
  [arXiv:hep-lat/0304018].


\bibitem{Gockeler:2003jfa}
  M.~Gockeler, R.~Horsley, D.~Pleiter, P.~E.~L.~Rakow, A.~Schafer, G.~Schierholz and W.~Schroers
                  [QCDSF Collaboration],
  Phys.\ Rev.\ Lett.\  {\bf 92}, 042002 (2004)
  [arXiv:hep-ph/0304249].

\bibitem{Ossmann:2004bp}
  J.~Ossmann, M.~V.~Polyakov, P.~Schweitzer, D.~Urbano and K.~Goeke,
  Phys.\ Rev.\  D {\bf 71}, 034011 (2005)
  [arXiv:hep-ph/0411172].

\bibitem{Guidal:2004nd}
  M.~Guidal, M.~V.~Polyakov, A.~V.~Radyushkin and M.~Vanderhaeghen,
  Phys.\ Rev.\  D {\bf 72}, 054013 (2005)
  [arXiv:hep-ph/0410251].

\bibitem{Diehl:2004cx}
  M.~Diehl, T.~Feldmann, R.~Jakob and P.~Kroll,
  Eur.\ Phys.\ J.\  C {\bf 39}, 1 (2005)
  [arXiv:hep-ph/0408173].

\bibitem{Wakamatsu:2005vk}
  M.~Wakamatsu and H.~Tsujimoto,
  Phys.\ Rev.\  D {\bf 71}, 074001 (2005)
  [arXiv:hep-ph/0502030].

\bibitem{Wakamatsu:2006dy}
  M.~Wakamatsu and Y.~Nakakoji,
  Phys.\ Rev.\  D {\bf 74}, 054006 (2006)
  [arXiv:hep-ph/0605279].






\bibitem{Soper:1976jc}
  D.~E.~Soper,
  Phys.\ Rev.\  D {\bf 15}, 1141 (1977).


\end{thebibliography}
\end{document}